\documentclass[twocolumn,aps,prl,showpacs,amsfonts]{revtex4}

\usepackage{graphicx}
\usepackage{amsmath}
\usepackage{amssymb}

\begin{document}
\title{Experimentally testable state-independent quantum contextuality}
\author{Ad\'{a}n Cabello$^{1}$}
\email{adan@us.es}
\affiliation{$^{1}$Departamento de F\'{\i}sica
Aplicada II, Universidad de Sevilla, E-41012 Sevilla, Spain}
\date{\today}



\begin{abstract}
We show that there are Bell-type inequalities for noncontextual
theories that are violated by {\em any} quantum state. One of these
inequalities between the correlations of compatible measurements is
particularly suitable for testing this {\em state-independent}
violation in an experiment.
\end{abstract}



\maketitle




Because of the lack of spacelike separation between one observer's
choice and the other observer's outcome, the immense majority of the
experimental violations of Bell inequalities does not prove quantum
nonlocality, but just quantum contextuality. Bell inequalities can
only be violated by entangled states. However, in principle,
Bell-type inequalities for noncontextual theories might be violated
by {\em any} quantum state.

Bell's theorem states that no theory of local hidden variables can
reproduce quantum mechanics (QM) \cite{Bell64}. It is proven either
by the violation of a Bell inequality \cite{Bell64, CHSH69} or by a
logical contradiction between the LHV predictions and those of QM
\cite{GHZ89}. Bell inequalities have some advantages. They are
independent of QM, testable in experiments, and have applications in
communication complexity \cite{BZPZ04}, entanglement detection
\cite{HGBL05}, security of key distribution \cite{AGM06}, and
quantum state discrimination \cite{SKLWZW08}. Any proof of Bell's
theorem is state-dependent: it is valid for some states but not for
others.

Local hidden variable theories are a special type of noncontextual
hidden variable (NCHV) theories, defined as those where the
expectation value of an observable $A$ is the same whether $A$ is
measured with a compatible observable $B$, or with a compatible
observable $C$, even though $B$ and $C$ are incompatible. The
Kochen-Specker (KS) theorem states that no NCHV theory can reproduce
QM \cite{Specker60, Bell66, KS67}. The KS theorem is proven by a
logical contradiction \cite{KS67, Peres90, Mermin90, Peres91,
Mermin93, Peres93, CEG96}. These proofs apply to systems described
by Hilbert spaces of dimension $d \ge 3$ and are state-independent
(i.e., valid for any state). Quantum contextuality is related to
quantum error correction \cite{DP97}, random access codes
\cite{Galvao02}, quantum key distribution \cite{Nagata05},
one-location quantum games \cite{AV08}, and entanglement detection
between internal degrees of freedom.

The differences between the proofs of Bell's and the KS theorems
lead to the question of what is the connection between them. It has
been shown \cite{HR83, Cabello01a, Cabello01b} that any proof of the
KS theorem can be converted into a proof of impossibility of
``elements of reality'' \cite{EPR35}. Some proofs of the KS theorem
can be converted into logical proofs of Bell's theorem
\cite{Mermin90, Mermin93} which can be translated into Bell
inequalities \cite{Mermin90a}.

The differences between the proofs are also in the heart of the
controversy on whether experimental tests of the KS theorem make
sense \cite{CG98} or are even possible if the finite precision of
measurements is taken into account \cite{Meyer99, Kent99, CK00,
SZWZ00, HKSS01, Appleby00b, SBZ01, Larsson02, Cabello02, HLZPG03}.

As a result of these debates, two types of inequalities to test
quantum contextuality have been proposed. On one hand, there are
``KS inequalities'' \cite{SBZ01, Larsson02}, which are based on the
assumption of contextuality and on some QM predictions, and
therefore are not independent of QM. On the other hand, there are
inequalities that are based {\em only} on the assumption of
noncontextuality, in the same way that the Bell inequalities are
based only on the assumption of locality. These inequalities are
independent of QM and testable in experiments. There are recent
proposals for testing inequalities of this type in different
physical systems \cite{CFRH08, KCBS08, Nambu08}. However, the fact
that all these inequalities are state-dependent, while the proofs of
the KS theorem are state-independent, has been recently described as
``a drawback'' \cite{KCBS08}. A natural question is the following:
Given a physical system described in QM by a Hilbert space of
dimension $d$, is it possible to derive experimentally testable
inequalities using {\em only} the assumption of noncontextuality,
such that {\em any} quantum state violates them?

We describe the first inequalities of this type. Each of them is
valid for a value of $d$, and all of them share a curious property.
Then, we will discuss how these inequalities may be tested, and
whether the state-independent violation predicted by QM can be
observed in actual experiments.




{\em First inequality.---}Suppose that $A_{ij}$ is an observable
with two possible results: $-1$ or $+1$, and two observables
$A_{ij}$ and $A_{kl}$ are compatible if they share a subindex (i.e.,
$i=k$, or $i=l$, or $j=k$, or $j=l$). When we prepare an ensemble of
systems and measure 4 compatible observables $A_{ij}$, $A_{ik}$,
$A_{il}$, and $A_{im}$ in each system, $\langle A_{ij} A_{ik} A_{il}
A_{im} \rangle$ denotes the average of the products of their
results. In any theory of NCHV in which the observables $A_{ij}$
have definite results, the following inequality must be satisfied:
\begin{align}
&-\langle A_{12} A_{16} A_{17} A_{18} \rangle-\langle A_{12} A_{23}
A_{28} A_{29}\rangle - \langle A_{23} A_{34} A_{37} A_{39} \rangle
\nonumber \\ &- \langle A_{34} A_{45} A_{47} A_{48} \rangle -
\langle A_{45} A_{56} A_{58} A_{59} \rangle - \langle A_{16} A_{56}
A_{67} A_{69} \rangle \nonumber \\ &- \langle A_{17} A_{37} A_{47}
A_{67} \rangle -\langle A_{18} A_{28} A_{48} A_{58} \rangle
\nonumber
\\ &-
\langle A_{29} A_{39} A_{59} A_{69} \rangle \le 7. \label{first}
\end{align}
This can be proven as follows. We define $\alpha = - A_{12} A_{16}
A_{17} A_{18}- \ldots - A_{29} A_{39} A_{59} A_{69}$. If we generate
all the $2^{18}$ possible values of $\alpha$, we will find that
$\alpha=7$ is the maximum. Therefore, if we can measure $\alpha$ on
different systems, the average satisfies $\langle \alpha \rangle \le
7$. We cannot measure $\alpha$ on a single system, because $\alpha$
contains incompatible observables. However, since we are assuming
that each $A_{ij}$ would give the same result in any context, we can
measure subsets of compatible observables on different subensembles
prepared in the same state, and then inequality (\ref{first}) is
valid for the averages over each subsensemble. This derivation is
similar to a standard derivation of a Bell inequality. The only
difference is that in a Bell inequality we assume that the result of
a measurement of $A_{12}$ is independent of spacelike separated
measurements, while here we assume that it is independent of
compatible measurements.


\begin{figure}
\centerline{\includegraphics[width=8.5cm]{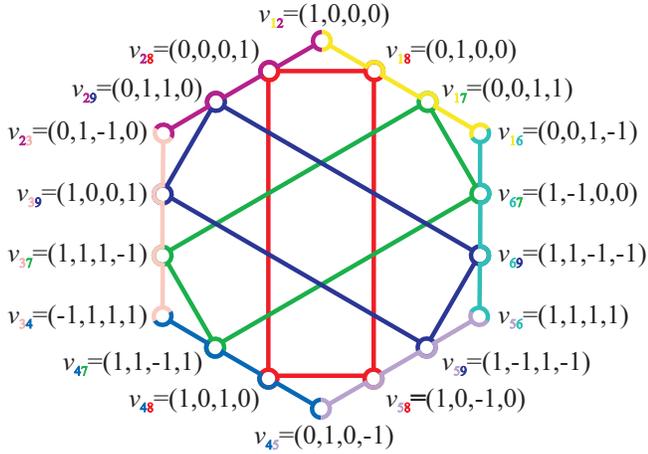}}
\caption{\label{Fig1} Each dot represents a unit-vector $v_{ij}$.
Each of the $6$ sides of the regular hexagon and each of the $3$
rectangles contains only orthogonal vectors. Note that, for
clarity's sake, most labels have no unit length.}
\end{figure}


Now consider a physical system described by a Hilbert space of
dimension $d=4$ (e.g., two qubits or a single spin-$3/2$ particle),
and the observables represented by the operators
\begin{equation}
A_{ij}=2 |v_{ij}\rangle\langle v_{ij}|-\openone, \label{observable}
\end{equation}
where $v_{ij}$ is a unit vector and $\openone$ denotes the identity.
Each observable $A_{ij}$ has two possible results: $-1$ or $+1$. If
$v_{ij}$ is orthogonal to $v_{ik}$, then $A_{ij}$ and $A_{ik}$ are
compatible. Therefore, $4$ orthogonal vectors define $4$ compatible
observables. $18$ vectors $v_{ij}$ with the the orthogonality
relations assumed in inequality (\ref{first}) are presented in
Fig.~\ref{Fig1}.

Let us prove that, for $d=4$, QM violates (\ref{first}) {\em for any
state}. According to QM, if one measures on the same system 4
compatible observables $A_{ij}$ corresponding to 4 orthogonal
vectors $v_{ij}$, the product of their $4$ results will always be
$-1$, because $A_{ij} A_{ik} A_{il} A_{im} = -\openone$. Therefore,
using the vectors of Fig. \ref{Fig1}, QM predicts that the
experimental value of the left-hand side of inequality (\ref{first})
must be $9$ {\em in any state}, which is clearly beyond the bound
for any description based on noncontextual hidden variables.




{\em Relation to previous results.---}The $18$ vectors in Fig.
\ref{Fig1} have also be used for a proof of the KS theorem
\cite{CEG96}, in which it is assumed: (I) that the observables
represented by the projectors $|v_{ij}\rangle \langle v_{ij}|$ have
noncontextual results $0$ or $1$, and (II) that the results of 4
compatible projectors are one 1 and 3 zeroes (a QM prediction). A
simple parity argument proves that it is impossible to assign values
satisfying both (I) and (II): There are 9 (an odd number) complete
sets of projectors, while each projector appears in two (an even
number) of them \cite{CEG96}. Here we have used the 18 vectors of
Fig. \ref{Fig1} for a different purpose: (\ref{first}) is an
experimentally testable inequality, not a proof by contradiction.

Other interesting relation of inequality (\ref{first}) to previous
results is the following. In Ref. \cite{KCBS08} there is a
state-dependent inequality for testing NCHV theories in systems of
$d = 3$ (e.g., spin-1 particles). Inequality (5) in Ref.
\cite{KCBS08} can be expressed as
\begin{eqnarray}
-\langle A_{12} A_{18} \rangle-\langle A_{12} A_{23}\rangle -
\langle A_{23} A_{34} \rangle - \langle A_{34} A_{48} \rangle
\nonumber \\ - \langle A_{18} A_{48}\rangle \le 3. \label{KCBS}
\end{eqnarray}
Using the observables $A_{ij}$ defined before, it is easy to see
that, for the state $(\cos{0.3},\sin{0.3})\otimes
(\cos{0.7},-\sin{0.7})$, the left-hand side of (\ref{KCBS}) is
$3.6$. Therefore, this two-qubit state violates inequality
(\ref{KCBS}). However, other states, e.g. the state $(1,0) \otimes
(1,0)$, do not violate it. The interesting observation is that
inequality (\ref{KCBS}) is a particular case of inequality
(\ref{first}): It can be obtained from (\ref{first}) by replacing 13
out of the 18 observables $A_{ij}$ with identities. While
(\ref{KCBS}) is a state-dependent inequality for systems of $d = 3$
and $d = 4$ \cite{KCBS08}, (\ref{first}) is a state-independent
inequality for systems of $d = 4$. Note that (\ref{first}) contains
several inequalities like (\ref{KCBS}).




{\em Second inequality.---}Suppose that $P_{ij}$, with $i \in
\{1,2,3\}$ and $j \in \{4,5,6\}$, is an observable with two possible
results: $-1$ or $+1$, and two observables $P_{ij}$ and $P_{kl}$ are
compatible if they share a subindex. Using the method described
before, it can be easily proved that any NCHV theory in which the
observables $P_{ij}$ have definite results satisfies the following
inequality:
\begin{eqnarray}
\langle P_{14} P_{15} P_{16} \rangle + \langle P_{24} P_{25} P_{26}
\rangle + \langle P_{34} P_{35} P_{36} \rangle +\langle P_{14}
P_{24} P_{34} \rangle \nonumber \\ + \langle P_{15} P_{25} P_{35}
\rangle - \langle P_{16} P_{26} P_{36} \rangle \le 4. \label{second}
\end{eqnarray}
However, if we consider a two-qubit system and choose the following
observables:
\begin{subequations}
\begin{align}
&P_{14}=Z_1,\;\;\;P_{15}=Z_2,\;\;\;
P_{16}=Z_1 \otimes Z_2, \label{Peresobservables1} \\
&P_{24}=X_2,\;\;\;P_{25}=X_1,\;\;\;
P_{26}=X_1 \otimes X_2, \\
&P_{34}=Z_1 \otimes X_2,\;\;\;P_{35}=X_1 \otimes
Z_2,\;\;\;P_{36}=Y_1 \otimes Y_2, \label{Peresobservablesn}
\end{align}
\end{subequations}
where, e.g., $Z_1$ denotes $\sigma_z^{(1)}$, the Pauli matrix $Z$ of
qubit~1, then, according to QM, the left-hand side of (\ref{second})
must be $6$, since $P_{14} P_{15} P_{16}=P_{24} P_{25} P_{26}=P_{34}
P_{35} P_{36}=P_{14} P_{24} P_{34}=P_{15} P_{25} P_{35}=-P_{16}
P_{26} P_{36}=\openone$. Therefore, QM violates inequality
(\ref{second}) for {\em any} two-qubit state.




{\em Relation to previous results.---}The observables
(\ref{Peresobservables1})--(\ref{Peresobservablesn}) have been used
in the proof of the KS theorem for two-qubit systems proposed by
Peres and Mermin \cite{Peres90, Mermin90, Mermin93}. This proof is
also based on a parity argument. Again, a testable inequality is
connected to a KS proof by contradiction based on a parity argument.

Other interesting connections of inequality (\ref{second}) to some
recent results are the following. In Ref. \cite{CFRH08} there is a
state-dependent inequality for testing quantum contextuality in
two-qubit systems. Inequality (4) in Ref. \cite{CFRH08} can be
expressed as
\begin{eqnarray}
- \langle P_{14} P_{15} \rangle - \langle P_{24} P_{25} \rangle -
\langle P_{34} P_{35} \rangle + \langle P_{14} P_{24} P_{34} \rangle
\nonumber \\ + \langle P_{15} P_{25} P_{35} \rangle \le 3.
\label{CFRH}
\end{eqnarray}
According to QM, for the singlet state, the left-hand side of
inequality (\ref{CFRH}) is 5 \cite{CFRH08}. What is interesting is
that inequality (\ref{CFRH}) is a particular case of inequality
(\ref{second}), when $P_{16}$, $P_{26}$, and $P_{36}$ are replaced
with $-\openone$.

Moreover, the recent proposal for testing quantum contextuality in
two-qubit systems in Ref. \cite{Nambu08} can be reformulated as the
following inequality:
\begin{eqnarray}
\langle P_{14} P_{15} P_{16} \rangle + \langle P_{24} P_{25} P_{26}
\rangle + \langle P_{34} P_{35} \rangle + \langle P_{14} P_{24}
P_{34} \rangle \nonumber \\ + \langle P_{15} P_{25} P_{35} \rangle -
\langle P_{16} P_{26} \rangle \le 4, \label{Nambu}
\end{eqnarray}
which is maximally violated by the product state
$|\sigma_y^{(1)}=+1\rangle \otimes |\sigma_y^{(2)}=+1\rangle$.
Inequality (\ref{Nambu}) is a particular case of inequality
(\ref{second}) when $P_{36}$ is replaced with $\openone$. The fact
that a product state violates inequality (\ref{Nambu}) is not
surprising, since {\em any} state violates inequality
(\ref{second}).

Finally, if $P_{15}=P_{25}=P_{34}=P_{35}=P_{36}=\openone$, then
inequality (\ref{second}) becomes
\begin{equation}
\langle P_{14} P_{16} \rangle + \langle P_{24} P_{26} \rangle
+\langle P_{14} P_{24} \rangle - \langle P_{16} P_{26} \rangle \le
2,
\end{equation}
which has the same structure of the Clauser-Horne-Shimony-Holt Bell
inequality \cite{CHSH69}.




{\em Third inequality.---}Suppose that the $4+2 n$ observables
${\cal A}_1,\ldots, {\cal A}_4, {\cal B}_1,\ldots, {\cal B}_n, {\cal
C}_1,\ldots, {\cal C}_n$, with $n$ (odd) $\ge 3$, have only two
possible results: $-1$ or $+1$. Assuming that each of the following
averages contains only compatible observables, using the method
described before, it can be easily seen that any NCHV theory
satisfies the following inequality:
\begin{eqnarray}
\left\langle {\cal A}_1 {\cal B}_1 {\cal B}_2 \prod_{i=3}^n {\cal
B}_i \right\rangle + \left\langle {\cal A}_2 {\cal B}_1 {\cal C}_2
\prod_{i=3}^n {\cal C}_i \right\rangle \nonumber \\ + \left\langle
{\cal A}_3 {\cal C}_1 {\cal B}_2 \prod_{i=3}^n {\cal C}_i
\right\rangle + \left\langle {\cal A}_4 {\cal C}_1 {\cal C}_2
\prod_{i=3}^n {\cal B}_i \right\rangle \nonumber
\\ - \langle {\cal A}_1 {\cal A}_2 {\cal A}_3 {\cal A}_4 \rangle \le 3.
\label{third}
\end{eqnarray}
However, if we consider an $n$-qubit system, with $n$ (odd) $\ge 3$,
and choose the following observables:
\begin{subequations}
\begin{align}
& {\cal A}_1=Z_1 \otimes Z_2 \otimes Z_3 \otimes \ldots \otimes Z_n, \label{merminobservables1}\\
& {\cal A}_2=Z_1 \otimes X_2 \otimes X_3 \otimes \ldots \otimes X_n, \\
& {\cal A}_3=X_1 \otimes Z_2 \otimes X_3 \otimes \ldots \otimes X_n, \\
& {\cal A}_4=X_1 \otimes X_2 \otimes Z_3 \otimes \ldots \otimes Z_n, \\
& {\cal B}_i = Z_i,\\
& {\cal C}_i = X_i, \label{merminobservablesn}
\end{align}
\end{subequations}
then, according to QM, the left-hand side of inequality
(\ref{third}) must be $5$, since, ${\cal A}_1 {\cal B}_1 {\cal B}_2
\prod_{i=3}^n {\cal B}_i = {\cal A}_2 {\cal B}_1 {\cal C}_2
\prod_{i=3}^n {\cal C}_i = {\cal A}_3 {\cal C}_1 {\cal B}_2
\prod_{i=3}^n {\cal C}_i = {\cal A}_4 {\cal C}_1 {\cal C}_2
\prod_{i=3}^n {\cal B}_i = - {\cal A}_1 {\cal A}_2 {\cal A}_3 {\cal
A}_4 = \openone$. Therefore, QM violates inequality (\ref{third})
for {\em any} $n$-qubit state with $n$ (odd) $\ge 3$.




{\em Relation to previous results.---}For $n=3$, the observables
(\ref{merminobservables1})--(\ref{merminobservablesn}) have been
used in a proof of the KS theorem for 3-qubit systems proposed by
Mermin \cite{Mermin90, Mermin93}. Again, Mermin's KS proof is a
proof by contradiction based on a parity argument.

On the other hand, taking ${\cal A}_1 = {\cal A}_2 = {\cal A}_3 = -
{\cal A}_4 = -\openone$, inequality (\ref{third}) becomes
\begin{eqnarray}
\left\langle {\cal B}_1 {\cal B}_2 \prod_{i=3}^n {\cal B}_i
\right\rangle + \left\langle {\cal B}_1 {\cal C}_2 \prod_{i=3}^n
{\cal C}_i \right\rangle + \left\langle {\cal C}_1 {\cal B}_2
\prod_{i=3}^n {\cal C}_i \right\rangle \nonumber \\ - \left\langle
{\cal C}_1 {\cal C}_2 \prod_{i=3}^n {\cal B}_i \right\rangle \le 2.
\label{Mermin}
\end{eqnarray}
What is interesting is that inequality (\ref{Mermin}) is not only a
state-dependent inequality to test quantum contextuality, but also a
Bell inequality. Indeed, for $n=3$, inequality (\ref{Mermin}) is the
$3$-party Bell inequality discovered by Mermin \cite{Mermin90a}. For
higher values of $n$, inequality (\ref{Mermin}) is not the Mermin
inequality \cite{Mermin90a}, but a new Bell inequality.




{\em Experimental violation.---}Observing the state-independent
violation predicted by QM in an actual experiment is a major
challenge for the near future. Inequality (\ref{second}) seems
particularly suitable for that purpose, since most of the
requirements for the experiment have been addressed, at least in the
case where the physical system is a two-qubit system consisting of
the spatial and spin components of a single neutron \cite{CFRH08}.
Other possibility is using the polarization and path degrees of
freedom of a single photon \cite{HLZPG03}. Using the polarization of
two photons, as proposed by \cite{Nambu08}, requires further
investigation in order to fulfill all the requirements of the
experiment.

To test inequality (\ref{second}), one has to prepare a specific
two-qubit quantum state (e.g., a maximally entangled state),
measure, e.g., $P_{14}$, $P_{15}$, and $P_{16}$, then prepare
another system in the same state and measure, e.g., $P_{24}$,
$P_{25}$, and $P_{26}$, and repeat these measurements many times,
until enough data has been obtained to calculate the 6 mean values
in (\ref{second}) and the experimental value of the Bell operator
for this state.

Then, one has to repeat the experiment with different states (e.g.,
a partially entangled state, a product state, and a maximally mixed
state). The violation predicted by QM is the same for every state.

There are two requirements for these experiments to be considered
legitimate state-independent tests of quantum contextuality: (a) The
experimental apparatus used for measuring, e.g., $P_{14}$ must be
the same when $P_{14}$ is measured together with $P_{15}$ and
$P_{16}$, and when it is measured together with $P_{24}$ and
$P_{34}$, and must be the same for any state. (b) Every observable
must be measured in different contexts. For a more detailed
discussion, see \cite{CFRH08}.




{\em Conclusions.---}We have introduced 3 experimentally testable
inequalities valid for any NCHV theory and violated by any quantum
state. They combine the most celebrated properties of the Bell
inequalities, independence of QM and experimental testability, with
state independence, the most celebrated property of the KS theorem.
One of these inequalities seems particularly suitable to
experimentally test the state-independent violation predicted by QM.

The connection of these inequalities to previous proofs of the KS
theorem and previous state-dependent inequalities gives a new
insight on the relationship between the two main theorems of
impossibility of hidden variables in QM. Each of the 3 introduced
state-independent inequalities is related to a proof of the KS
theorem based on a parity argument. An open question is whether
similar state-independent inequalities can be developed for physical
systems where no proofs of the KS theorem based on a parity argument
are known. Specifically, an interesting open problem is finding a
state-independent inequality based only on the assumption of
noncontextuality for the case $d=3$.


\begin{acknowledgments}
The author acknowledges support from projects No. P06-FQM-02243 and
No. FIS2008-05596.
\end{acknowledgments}



\begin{thebibliography}{99}

\bibitem{Bell64}
J. S. Bell,
Physics (Long Island City, NY) {\bf 1}, 195 (1964).
%
\bibitem{CHSH69}
J. F. Clauser, M. A. Horne, A. Shimony, and R. A. Holt,
Phys. Rev. Lett. {\bf 23}, 880 (1969).

\bibitem{GHZ89}
D. M. Greenberger, M. A. Horne, and A. Zeilinger,
in {\em Bell's Theorem, Quantum Theory, and Conceptions of the
Universe}, edited by M. Kafatos (Kluwer Academic, Dordrecht, 1989),
p.~69.

\bibitem{BZPZ04}
\v{C}. Brukner, M. \.{Z}ukowski, J.-W. Pan, and A. Zeilinger,
Phys. Rev. Lett. {\bf 92}, 127901 (2004).

\bibitem{HGBL05}
P. Hyllus, O. G\"uhne, D. Bru\ss, and M. Lewenstein,
Phys. Rev. A {\bf 72}, 012321 (2005).

\bibitem{AGM06}
A. Ac\'{\i}n, N. Gisin, and L. Masanes,
Phys. Rev. Lett. {\bf 97}, 120405 (2006).

\bibitem{SKLWZW08}
C. Schmid, N. Kiesel, W. Laskowski, W. Wieczorek, M. \.{Z}ukowski,
and H. Weinfurter,
Phys. Rev. Lett. {\bf 100}, 200407 (2008).

\bibitem{Specker60}
E. P. Specker,
Dialectica {\bf 14}, 239 (1960).

\bibitem{Bell66}
J. S. Bell,
Rev. Mod. Phys. {\bf 38}, 447 (1966).

\bibitem{KS67}
S. Kochen and E. P. Specker,
J. Math. Mech. {\bf 17}, 59 (1967).


\bibitem{Peres90}
A. Peres,
Phys. Lett. A {\bf 151}, 107 (1990).
%
\bibitem{Mermin90}
N. D. Mermin,
Phys. Rev. Lett. {\bf 65}, 3373 (1990).


\bibitem{Peres91}
A. Peres,
J. Phys. A {\bf 24}, L175 (1991).

\bibitem{Mermin93}
N. D. Mermin,
Rev. Mod. Phys. {\bf 65}, 803 (1993).

\bibitem{Peres93}
A. Peres,
{\em Quantum Theory: Concepts and Methods}
(Kluwer, Dordrecht, 1993).

\bibitem{CEG96}
A. Cabello, J. M. Estebaranz, and G. Garc\'{\i}a-Alcaine,
Phys. Lett. A {\bf 212}, 183 (1996).


\bibitem{DP97}
D. P. DiVincenzo and A. Peres,
Phys. Rev. A {\bf 55}, 4089 (1997).

\bibitem{Galvao02}
E. F. Galv\~{a}o,
Ph.\ D. thesis, Oxford University, 2002.

\bibitem{Nagata05}
K. Nagata,
Phys. Rev. A {\bf 72}, 012325 (2005).

\bibitem{AV08}
N. Aharon and L. Vaidman,
Phys. Rev. A {\bf 77}, 052310 (2008).


\bibitem{HR83}
P. Heywood and M. L. G. Redhead,
Found. Phys. {\bf 13}, 481 (1983).

\bibitem{Cabello01a}
A. Cabello,
Phys. Rev. Lett. {\bf 86}, 1911 (2001).

\bibitem{Cabello01b}
A. Cabello,
Phys. Rev. Lett. {\bf 87}, 010403 (2001).

\bibitem{EPR35}
A. Einstein, B. Podolsky, and N. Rosen,
Phys. Rev. {\bf 47}, 777 (1935).

\bibitem{Mermin90a}
N. D. Mermin,
Phys. Rev. Lett. {\bf 65}, 1838 (1990).


\bibitem{CG98}
A. Cabello and G. Garc\'{\i}a-Alcaine,
Phys. Rev. Lett. {\bf 80}, 1797 (1998).

\bibitem{Meyer99}
D. A. Meyer,
Phys. Rev. Lett. {\bf 83}, 3751 (1999).

\bibitem{Kent99}
A. Kent,
Phys. Rev. Lett. {\bf 83}, 3755 (1999).

\bibitem{CK00}
R. Clifton and A. Kent,
Proc. R. Soc. London, Ser. A {\bf 456}, 2101 (2000).

\bibitem{SZWZ00}
C. Simon, M. \.{Z}ukowski, H. Weinfurter, and A. Zeilinger,
Phys. Rev. Lett. {\bf 85}, 1783 (2000).

\bibitem{HKSS01}
H. Havlicek, G. Krenn, J. Summhammer, and K. Svozil,
J. Phys. A {\bf 34}, 3071 (2001).

\bibitem{Appleby00b}
D. M. Appleby,
Phys. Rev. A {\bf 65}, 022105 (2002).

\bibitem{SBZ01}
C. Simon, \v{C}. Brukner, and A. Zeilinger,
Phys. Rev. Lett. {\bf 86}, 4427 (2001).

\bibitem{Larsson02}
J.-\AA. Larsson,
Europhys. Lett. {\bf 58}, 799 (2002).

\bibitem{Cabello02}
A. Cabello,
Phys. Rev. A {\bf 65}, 052101 (2002).

\bibitem{HLZPG03}
Y.-F. Huang, C.-F. Li, Y.-S. Zhang, J.-W. Pan, and G.-C. Guo,
Phys. Rev. Lett. {\bf 90}, 250401 (2003).


\bibitem{CFRH08}
A. Cabello, S. Filipp, H. Rauch, and Y. Hasegawa,
Phys. Rev. Lett. {\bf 100}, 130404 (2008).

\bibitem{KCBS08}
A. A. Klyachko, M. A. Can, S. Binicio\u{g}lu, and A. S. Shumovsky,
Phys. Rev. Lett. {\bf 101}, 020403 (2008).

\bibitem{Nambu08}
Y. Nambu,
arXiv:0805.3398.


\end{thebibliography}
\end{document}